\documentclass[aps,prl,twocolumn,showpacs,amsmath,amssymb]{revtex4}
\usepackage{epsfig}
\usepackage{bm}
\def\be{\begin{equation}}
\def\ee{\end{equation}}

\begin{document}
\title{Transport Statistics of Bistable Systems}
\author{Andrew N. Jordan and Eugene V. Sukhorukov}
\affiliation{D\'epartement de Physique Th\'eorique, Universit\'e de Gen\`eve,
        CH-1211 Gen\`eve 4, Switzerland}
\date{June 11, 2004}

\begin{abstract}
We consider the transport statistics of classical bistable systems
driven by noise.  The stochastic path integral formalism is used to
investigate the dynamics and distribution of transmitted charge.
Switching rates between the two stable states are found from an
instanton calculation, leading to an effective two-state system on a
long time scale.  In the bistable current range, the telegraph noise
dominates the distribution, whose logarithm is found to be universally
described by a tilted ellipse.
\end{abstract}

\maketitle

Bistable systems are ubiquitous in nature.  They occur in solid state
 \cite{SS,SS2}, magnetic \cite{magnetic}, and
biological systems \cite{bio}.  The behavior of bistable systems may
be characterized by random switching between two stable states,
induced by either thermal fluctuations, or non-equilibrium noise.  The
presence of bistability is usually deduced by measuring the current
passing through the system.  The switching of the internal state of
the system alters the resistance of the sample, causing the measured
current to randomly switch between two distinct average values.  The
noise of this switching current is called random telegraph noise
\cite{machlup}, a process studied by Machlup already in the 1950s.  An
important feature of this process is that its zero frequency noise
power diverges as the switching rates tend to zero.

Fluctuations in bistable electronic devices, such as tunnel diodes,
were theoretically investigated by Landauer \cite{diode}, and later
by H\"anggi and Thomas \cite{Hanggi}.  Since then, bistable fluctuations
have been observed in microstructures by different experimental groups
\cite{RTNexps}.  The switching times were found to be well fit by an
exponential of the applied bias.  More recently, similar physics was
discovered in mesoscopic resonant tunneling wells near an instability
\cite{rtw}.
\begin{figure}[thb]
\begin{center}
\leavevmode
\psfig{file=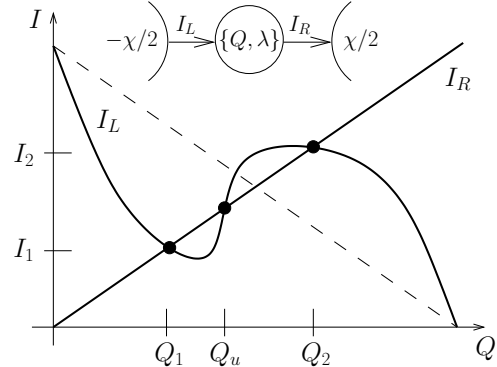,width=6.5cm}
\caption{A schematic of the average current flowing through the
left and right connector as a function of the charge in the central node.
While linear connectors (the dashed line) have only one crossing
point, nonlinear devices may have three crossing points, two stable, 
$Q_{1,2}$, and one unstable, $Q_u$.
Inset: The stochastic network.  The direction of current flow through
the network is indicated.}
\label{net}
\end{center}
\vspace{-5mm}
\end{figure}
In this Letter, we examine the properties of bistable systems from the
novel point of view of transport statistics.  This approach gives not
only the average current, and zero-frequency noise \cite{BB}, but all higher
current cumulants as well (also called full counting
statistics in mesoscopic physics \cite{FCS}).  
For many systems, high order cumulants may be computed via the cascade
method \cite{Nagaev}, which uses the idea that the noise itself has
noise \cite{Nagaev,cascade,reulet}, and that feedback effects of the lower 
order cumulants may be
taken into account in a perturbative manner \cite{us1,us2}.
In contrast, for unstable systems, non-linearities enter the transport
in an essentially {\it non-perturbative} fashion.  
Here the situation is reversed: All higher cumulants 
of the noise sources must be accounted for to correctly find the switching 
rates which govern even the average current.

In order to successfully describe the physics of such an instability,
we apply the Stochastic Path Integral (SPI), a non-perturbative method
that was recently introduced \cite{us1,us2} to evaluate the
statistical fluctuations in a classical stochastic network.  The network is
composed of nodes containing effectively continuous charge
representing any conserved quantity. The charge is transferred through
connectors whose isolated transport properties are given.  In order to
emphasize the universality of the results and focus on the essential
physics, we use an abstract description which can represent many of
the physical systems mentioned thus far, such as the tunneling diode,
or more general S or Z-type instabilities \cite{SS2}.

We now briefly describe our main results.  We let the average left
current have a common S-type nonlinear dependence on the charge on the
nodes (see Fig.\ \ref{net}), and demonstrate bistability.  The
connector noise induces switching between the stable states with the
rates $\Gamma_{1,2}$, found via an instanton calculation, Eq.\
(\ref{delta}), to be exponentially small with the instability charge
scale.  On the time scale of order $\Gamma_{1,2}^{-1}$ or longer, the
system can be viewed as being an effectively two-state system.  After
this point, results (\ref{E}-\ref{asymp2}) hold independently of how
the rates $\Gamma_{1,2}$ are obtained. The requirement of continuous
charge may be relaxed, and the results apply to any telegraph process
\cite{SS,SS2,magnetic,bio} with the rates as input parameters.  The
distribution of the transmitted charge is dominated by one of the
stable states in the extreme small or large current ranges.  In
between, the distribution has a comparatively flat region dominated by
the telegraph process, Eq.\ (\ref{asymp2}). Given $\Gamma_{1,2}$, it
is universal: The logarithm of the distribution is a tilted ellipse.

{\it The model.}--The bistable network we consider consists of one
``conserving'' node, $C$, storing the generalized charge $Q$.  A
typical non-equilibrium fluctuation decays on a time scale $\tau_C$.
This node is connected to two ``absorbing'' nodes, $L$
and $R$, which inject and absorb currents $\tilde I_L$ and $\tilde
I_R$ through the connectors between the nodes (see inset of Fig.\
\ref{net}).  The sources of noise are the currents $\tilde I_{L,R}$,
which are Markovian random variables (after a time scale $\tau_0 \ll
\tau_C$) with known statistics given by the current cumulant
generating functions, 
\be
H_{\alpha}(\lambda_{\alpha})=\sum_n(\lambda_{\alpha}^n/n!)\langle\langle
\tilde I_{\alpha}^n\rangle\rangle, \quad\alpha=L,R,
\label{generators}
\ee 
where the cumulants $\langle\langle \tilde
I_{\alpha}^n\rangle\rangle$ are functions of the charge in the
central node, $Q$ \cite{footnote0}.  
Charge conservation, $\dot Q=\tilde I_L-\tilde
I_R$, leads to the slow dynamics of $Q$ which couples back into the
dynamics of $\tilde I_{L,R}$ by changing the conditions for 
transport through the connectors, thus altering the statistics of the
net current $I=(\tilde I_L+\tilde I_R)/2$.  The cumulants
$\langle\langle {\cal Q}^n\rangle\rangle$ of the {\it transmitted
charge} ${\cal Q}(t)=\int_0^t dt' I(t')$ can be obtained with
diagrammatic perturbation theory \cite{Nagaev,us2} in the stable
regime by using the SPI formalism. However, the power of the SPI
method is that it also allows for a non-perturbative analysis of
unstable systems.

The SPI represents the time evolution 
of the system while monitoring all fluctuations \cite{us1,us2}:
\be
U[\chi,t]=\!\int {\cal D}Q{\cal D}\lambda
\exp\Big\{\int_0^t dt'[-\lambda\dot Q
+H(Q,\lambda, \chi)]\Big\},
\label{pi}
\ee
with fixed initial charge $Q(0)$ and 
the final condition $\lambda(t)=0$. 
The generating function 
$S(\chi,t)$ for the cumulants of the transmitted charge  
$\langle\langle {\cal Q}^n\rangle\rangle = d_\chi^n S(\chi, t)
\vert_{\chi=0}$ is given by $S=\ln U[\chi, t]$.
The path integral (\ref{pi}) has a canonical form, 
with the ``Hamiltonian'' $H$ given by 
\be
H(Q, \lambda, \chi) = 
H_{L}(\lambda+\chi/2)+H_{R}(-\lambda + \chi/2), 
\label{htot}
\ee 
where $H_{L,R}$ are the current cumulant generators (\ref{generators}). 
The counting variable $\lambda$
is a Lagrange multiplier conserving the charge on
the node $C$. A large parameter, the number of elementary charges
participating in the transport, allows the saddle-point evaluation of
the SPI (\ref{pi}) leading to the canonical equations of
motion
\begin{equation}
\dot Q=\partial H/\partial \lambda,\quad
\dot\lambda=-\partial H/\partial Q.
\label{motion}
\end{equation}
The solution of these equations should be substituted 
into Eq.\ (\ref{pi}) in order to obtain the generating 
function $S$. 

\begin{figure}[thb]
\begin{center}
\leavevmode
\psfig{file=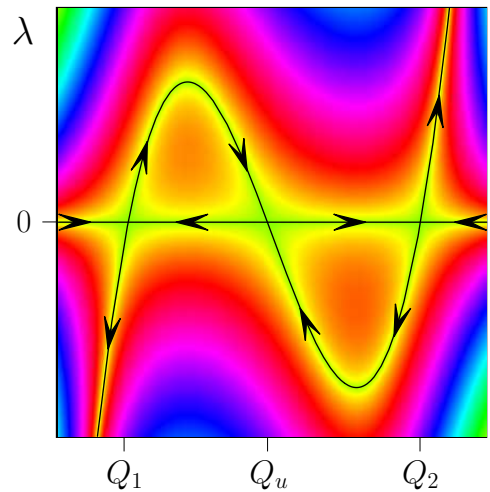,width=6.5cm}
\caption{The stability profile.  The Hamiltonian is given as a phase
space density plot.  The system has three fixed points, two stable at
$Q=Q_1, Q_2$ and one unstable at $Q=Q_u$.  For long time, the system
is forced to the zero energy lines in black, $\lambda_0,\lambda_{\rm
in}$.  A decay from the unstable state may happen with zero action,
but to hop from either stable state to the unstable state requires
finite action, the area underneath the curves shown.}
\label{cuts}
\end{center}
\vspace{-5mm}
\end{figure}

{\em The bistability.}--We first assume for simplicity 
that the sources $\tilde I_\alpha$, $\alpha=L,R$ are Gaussian 
with average current $I_\alpha(Q)=\langle\tilde I_\alpha\rangle$ 
and noise power
$F_\alpha(Q)=\langle\langle(\tilde I_\alpha)^2\rangle\rangle$,
and then generalize to arbitrary noise generators (\ref{generators}). 
Fixing $\chi=0$, we find that the unconditional 
dynamics of the system (where the transmitted charge is unmonitored) 
is governed by the Hamiltonian
\be
H_0 = (I_L-I_R)\lambda + (1/2)(F_L+F_R)\lambda^2.
\label{ham}
\ee
Therefore, the stationary state ${\dot Q} ={\dot \lambda} =0$ is
[according to Eq.\ (\ref{motion})] given by $\lambda=0$ and 
$I_L(Q) = I_R(Q)$, the conservation 
of average current.  In the ohmic regime, the
average current is a linear function of the
charge in the node (see Fig.\ \ref{net}), so the currents intersect 
only at one point.  However, if we allow a nonlinear I-V, 
topological distortions of the curves allow three
crossing points, giving three stationary solutions. 
We will now show that while the outer two intersections are stable 
to perturbations, the central intersection is unstable.

First, we note that because the Hamiltonian (\ref{ham}) does not
explicitly depend on time, it is an integral of motion: $H_0(Q,
\lambda)=E$, where the ``energy'' determines a trajectory in phase
space. On a time scale longer than the relaxation time, $t\gg\tau_C$,
the dissipative dynamics of the system projects a $Q$-distributed
initial state onto the zero energy lines, $H_0(Q,\lambda)=0$.
According to Eq.\ (\ref{ham}), there are at least two
such lines: The $\lambda_0$ line with $\lambda=0$, and the ``instanton''
line $\lambda_{\rm in}$, given by 
\be 
\lambda_{\rm in}(Q)=-2(I_L-I_R)/(F_L+F_R).
\label{instanton}
\ee
These lines, shown in Fig.\ \ref{cuts}, connect
the three stationary points, $Q_1$, $Q_u$, and $Q_2$.
From Eqs.\ (\ref{motion}), for the line $\lambda_0$ we have $\dot Q=I_L-I_R$, 
while for the $\lambda_{\rm in}$ line the sign is opposite, $\dot Q=I_R-I_L$. 
Then the simple investigation of Fig.\ \ref{net} shows the
propagation direction.  

According to Eq.\ (\ref{pi}), the action on the zero energy 
lines is given by $S=-\int\!\lambda\, dQ$. On the
$\lambda_0$ line, the action vanishes, which makes the dynamics along 
this line most probable. This dynamics describes the relaxation
of the initial state to the points $Q_1$ and $Q_2$ [see Fig.\ 
(\ref{cuts})], making them the stable stationary points of the system,
and $Q_u$ the unstable point.
The propagation along the $\lambda_{\rm in}$ line away
from stable points generates a non-zero action, and therefore
it has an exponentially small probability. 
Occasionally, the system switches between points $Q_1$
and $Q_2$ on the lines $\lambda_0$, $\lambda_{\rm in}$
with the action $S=-A_{1,2}$,
\be 
A_{1,2}= \int^{Q_u}_{Q_{1,2}} d Q\, \lambda_{\rm
in}(Q),
\label{act1}
\ee 
given
in Fig.\ \ref{cuts} by the area between the
lines $\lambda_0$ and $\lambda_{\rm in}$.

For an estimate, we consider the simplest (symmetric
cubic) current difference, $I_L-I_R=-\alpha Q[Q^2-(\Delta Q)^2]$, and
constant noise, $F_L+F_R = 2 {\bar F}$. For the action (\ref{act1})
with the instanton (\ref{instanton}), we
obtain $A_{1,2}= (\alpha/4 {\bar F})(\Delta Q)^4$.
By setting $A_{1,2}\sim 1$ and using Eqs.\ (\ref{motion})
we estimate the switching time $\tau_{\rm in}\sim(\Delta Q)^2/\bar F$,
a new time scale in the problem. The assumption of effectively continuous
charge demands that $\Delta Q\gg 1$, leading to $\tau_{\rm
in}\gg\tau_0$, so that the separation of time scales requirement is
satisfied. This also implies the instability is well developed.

The result (\ref{act1}) derived  above for Gaussian noise with 
the instanton (\ref{instanton}) holds for general noise 
with the $\lambda_{\rm in}$ line defined by the equation
\be
H_0(Q,\lambda)=H_L(\lambda)+H_R(-\lambda)=0.
\label{zero-energy}
\ee  
From now on we consider general noise $H_{L,R}$
\cite{footnote1}.

{\it The effective two-state Hamiltonian.}--We now proceed with the
instanton calculation \cite{kleinert}.  We consider a time interval
$\Delta t$ much longer than the switching time $\tau_{\rm in}$.
On this time scale, the system always relaxes to one of the stable points
$Q_{1,2}$, so that it can be considered as being effectively a two
state system where the evolution operator (\ref{pi})
becomes a $2\times2$ matrix. Starting from one of
the stable points, the system will switch to the other stable point
with the small transition probability $\Gamma_{1,2}\Delta t$ 
that is proportional to the time interval, 
because the attempts are uncorrelated.  
The switching rates are given by the instanton
($\lambda_{\rm in}$-line) contribution to the path integral, 
\be
\Gamma_{1,2} = \omega_{1,2}\exp(-A_{1,2}),
\label{delta}
\ee 
where the action $A_{1,2}$ is given by Eq.\ (\ref{act1}) \cite{footnote2}.
We note if the noise goes to zero, then
$\Gamma_{1,2}$ vanish exponentially.
The prefactor $\omega_{1,2}$ is the attempt rate,
 which is subdominant \cite{footnote3}.
Finally, using the smallness of the switching rates,
we can write a differential master equation for the evolution
operator, $\dot U=\hat H_0 U$, where the effective Hamiltonian 
matrix has components $\pm\Gamma_{1,2}$.  
The evolution operator describes the relaxation 
with the rate $\Gamma_S=\Gamma_1+\Gamma_2$ of any given initial state 
to the stationary state with constant occupation probabilities,
\be
P_1=\Gamma_2/\Gamma_S,\quad P_2=\Gamma_1/\Gamma_S.
\label{state}
\ee

Having described the bare (unconditional) dynamics of the
system, we now concentrate on the transport statistics.  A 
$\chi$ contribution to the Hamiltonian (\ref{ham})  deforms
the stability profile in Fig.\ \ref{cuts}. The stable stationary
points acquire a non-zero energy, and thus contribute to the diagonal
values in the full effective Hamiltonian $\hat H$.  The rates
$\Gamma_{1,2}$ will also depend on $\chi$.  However, the bistability
is only important for $\chi\propto\Gamma_{1,2}$, and therefore the
$\chi$ correction to $\Gamma_{1,2}$ is exponentially small and may be
neglected.  Using Eq.\ (\ref{htot}) to evaluate the contribution of
the stable points to the action, we arrive at the result 
\be \hat H =
\begin{pmatrix}
H_1(\chi) -\Gamma_1 & \Gamma_2 \cr
\noalign{\medskip}
\Gamma_1 & H_2(\chi)-\Gamma_2  
\end{pmatrix}.
\label{ham2}
\ee
The Hamiltonians $H_{1,2}(\chi)$ are the noise generators
at the stable points $Q_{1,2}$ 
with the net currents $I_{1,2}=I_L=I_R$ and noise power 
$F_{1,2}=(F_L G_R^2 + F_R G_L^2)/(G_L+G_R)^2$, where
$G_{L,R}$ are the differential conductances taken at $Q_{1,2}$.
High cumulants may be found perturbatively \cite{us2}. 

{\it The transport statistics.}-- The Hamiltonian (\ref{ham2})
determines the conditional dynamics of the effective two-state system
on the time scale $t\gg\tau_{\rm in}$ \cite{footnote4}.  
It describes the relaxation of
the system to the new {\it conditional} stationary state. In an
experiment on the bistable system, if the averaging time of the
measurement is much shorter then the relaxation time
$\tau_S=(\Gamma_1+\Gamma_2)^{-1}$, the output of the measurement
device will only reveal the noise properties of one of the states. On
the time scale $t\gg\tau_S$ the system switches many times and loses
its correlation with the initial state. As a result, the transport
becomes again Markovian, $S(\chi)=tH(\chi)$.
The generating function $H(\chi)$ is given by the largest
eigenvalue of the Hamiltonian (\ref{ham2}),
\begin{eqnarray}
H(\chi)&=&(1/2)\sum_{n=1,2}(H_n-\Gamma_n)\nonumber \\
&+& \sqrt{\left(H_2-H_1 
-\Delta\Gamma\right)^2/4+ \Gamma_1\Gamma_2},\quad
 \label{E}
\end{eqnarray}
where $\Delta \Gamma = \Gamma_2-\Gamma_1$.
\begin{figure}[thb]
\begin{center}
\leavevmode
\psfig{file=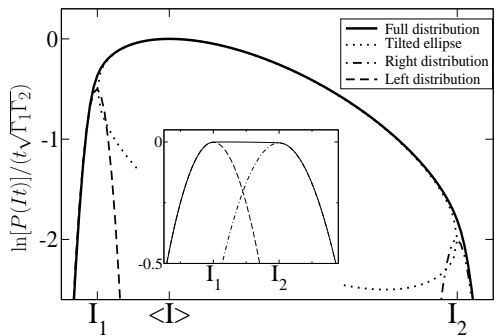,width=6.5cm}
\caption{
The log-distribution of transmitted charge, divided by $t \sqrt{\Gamma_1
\Gamma_2}$. Within the bistable range, the distribution is well fit by
a tilted ellipse, connecting to the distribution dominated by state 1
or 2. The average current is denoted by $\langle I \rangle$. 
Inset: The same distribution, $\ln P[It]/t$, showing the comparatively flat
bistable distribution with the left and right asymptotics.
$\Gamma_2/\Gamma_1=4,\, F_{1,2}=(1, 1.5) \times 10^{-2} \langle\langle
I^2\rangle\rangle$.}
\label{probsblow}
\end{center}
\vspace{-5mm}
\end{figure}
We immediately see that the distribution is properly
normalized because $H(0)=0$. The average current is given by
$\langle I\rangle=H'(0)=\sum_{n=1,2}I_n P_n$,
and the noise power $\langle\langle I^2\rangle\rangle=H''(0)$ by
\be
\langle\langle I^2\rangle \rangle
= \sum_{n=1,2}F_nP_n + 2 (\Delta I)^2\Gamma_1\Gamma_2/\Gamma_S^3,
\label{result1}
\ee where $\Delta I=I_2-I_1$, and the stationary probabilities $P_n$
are given in Eq.\ (\ref{state}).  The first term is the weighted noise
of the stationary states, and the second term is the well-known result
for zero-frequency telegraph noise \cite{machlup} which dominates the
first term because the rates are small. The dominant contribution to
the third cumulant is $\langle\langle I^3\rangle\rangle=6(\Delta
I)^3\Gamma_1\Gamma_2 \Delta\Gamma/\Gamma_S^5$.  The slow switching
rates cause all cumulants to be parametrically large, making 
bistable systems a promising candidate for an experimental measurement 
of the transport statistics.

We now Fourier transform $U(i \chi)$ using the stationary phase
approximation to obtain the probability to transmit charge ${\cal
Q}=It$ through the system in time $t$.  
Outside the bistability region, $I<I_1$ or $I>I_2$
the generator in Eq.\ (\ref{E}) may be replaced with $H_1-\Gamma_1$
or $H_2-\Gamma_2$, so that the extreme value statistics 
is dominated by one of the stable states. 
Within the bistable region, $I_1<I<I_2$,  
the probability ${\cal P}_n(I)$ of occupying 
state $n=1,2$ under the condition that a charge $It$ is
transmitted is given by 
\be {\cal P}_{1,2}(I) = {\cal G}_{2,1}/({\cal G}_1+{\cal G}_2),\quad {\cal
G}_n=\sqrt{\Gamma_n|I-I_n|},
\label{cp}
\ee
which generalizes (\ref{state}). The probability distribution 
of transmitted charge is given by
\be
\ln P(It)/t=-({\cal G}_1-{\cal G}_2)^2/\Delta I
\label{asymp2}
\ee
and is determined by the telegraph process.
The logarithm of the exact
distribution (divided by the rates) is shown in Fig.\ 3
inside the bistable region and is well fit by a tilted
ellipse, while the inset compares left and right asymptotics to the
full distribution.  

In conclusion, we have discussed a general model of a noise driven
bistable system.  We have calculated the switching rates
(\ref{delta}), transport statistics (\ref{E}) and (\ref{asymp2}), and
conditional occupation probabilities (\ref{cp}).  The higher cumulants
in Eq.~(\ref{generators}) determine the instanton lines in Fig. 2 [see
Eq.\ (\ref{zero-energy})], and thus make an exponentially large
contribution to the rates $\Gamma_{1,2}$ \cite{Hanggi,footnote1}.
Away from the bistable region, the probability tails decay with a
non-Gaussian distribution determined by the isolated statistics
$H_{1,2}$ of the stable states.  Within the bistable current range,
the distribution is always an ellipse whose dimensions are fixed by
the rates alone, showing a robust universality.

This work was supported by the SNF, and by INTAS 
(project 0014, open call 2001).


\begin{thebibliography}{02}
\bibitem{footnote0} 
One implementation of this model is an electrical
circuit, with bias $V$, where the metallic leads are the absorbing
nodes, the connectors are a macroscopic resistor, $R$, with Nyquist
noise, and tunnel diode with Poissonian statistics, while the central
node represents a parallel capacitor, $C$.  The semiclassical nature
of the circuit and the separation of time scales, implying that
$eV>\hbar/RC$, justify the use of the classical theory \cite{us1,us2}.
\bibitem{footnote2}
A similar instanton calculation has recently been done 
with the SPI in a different physical context, see \cite{kamenev}.
\bibitem{footnote4}
This description of transport statistics is formally
similar to that of Ref. \cite{nazarov} in Coulomb blockade systems.
\end{thebibliography}
\end{document}